\documentclass{article}
\usepackage{LaThuileFPSpro}
\usepackage{graphicx}
\usepackage{mathrsfs}

\usepackage{subfigure}
\usepackage{amssymb}
\usepackage{afterpage}

\newcommand*{\etaphi}{$\eta$-$\phi$}
\newcommand*{\Jone}{jet\#$1$}
\newcommand*{\TR}{``transverse"}

\newcommand*{\delphi}{$\Delta\phi$}
\newcommand*{\UE}{``underlying event"}

\newcommand{\py}{{\sc pythia}}
\newcommand{\pt}{$p_{T}$}

\newcommand{\gevcc}{$\mbox{GeV}/c^2$}
\newcommand{\gevc}{$\mbox{GeV}/c$}

\begin{document}
\title{ 
RECENT QCD STUDIES AT THE TEVATRON
  }
\author{
  Robert Craig Group        \\
  {\em Fermilab, Batavia, IL, USA} \\
  {\em (On behalf of the CDF and D\O\ Collaborations)}
  }
\maketitle

\baselineskip=11.6pt

\begin{abstract}
  Since the beginning of Run II at the Fermilab Tevatron, the QCD physics groups
  of the CDF and D\O\ experiments have worked to reach unprecedented levels of
  precision for many QCD observables.  Thanks to the large dataset - over
  $3~\mbox{fb}^{-1}$ of integrated luminosity  recorded by each experiment -
  many important new measurements have recently been made public and
  will be summarized in this paper.      
\end{abstract}
\newpage
\section{Introduction}
The Tevatron collider at Fermilab provides collisions of protons with
anti-protons at a center of mass energy of $1.96$~TeV.  This is
currently the highest energy collider in the world.  The multipurpose
detectors of the CDF~\cite{CDF_detector} and D\O~\cite{D0_detector} experiments are exploiting the more than
$3~\mbox{fb}^{-1}$ of integrated luminosity provided by the Tevatron
in order to make important progress in constraining and
confirming the calculations made from quantum chromodynamics (QCD).  

Precise measurements of QCD observables in hadron-hadron collisions -
such as jet cross sections - constrain parton density functions (PDFs)
and confirm the predictive power of theory.  This results in a better
control of the standard QCD production calculations which are used to predict
major backgrounds for many important physical processes.  In addition, the
specific QCD processes which pose challenges to new physics searches
such as supersymmetry and Higgs production can be measured directly with
dedicated analyses. 

In this paper
some of the most recent measurements from the CDF and D\O\
collaborations will be reviewed.  These measurements will be split
into underlying event observables, jet cross sections, and boson plus
jet cross section measurements.  

\section{Hadronic Collisions and Underlying Event Observables}

\begin{figure}[htp]
  \begin{center}
  \includegraphics[width=8.5cm]{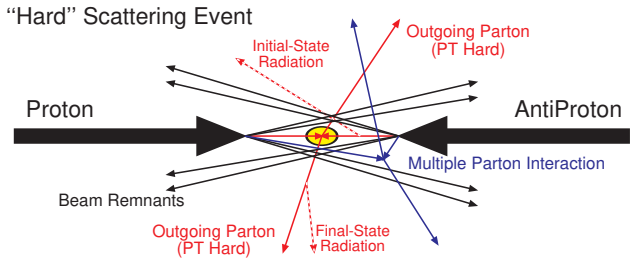}
  \end{center}
  \caption{\it
    Simple model for hadronic collisions.
    \label{fig:hard_scat} }
\end{figure}

A brief introduction into the structure of hadronic collisions is
useful as a motivation for jet definition.  Hadronic collisions may be
factorized into perturbative components (hard scattering and initial
and final state radiation) and non-perturbative components (beam
remnants and multiple parton interactions).  These components are
illustrated in the simple ``cartoon''
shown in figure~\ref{fig:hard_scat}.  This simple picture is similar
to the model used by a program like \py~\cite{Sjostrand:2000wi} to
generate hadronic collisions. 

\begin{figure}[htp]
  \begin{center}
  \includegraphics[width=6.5cm]{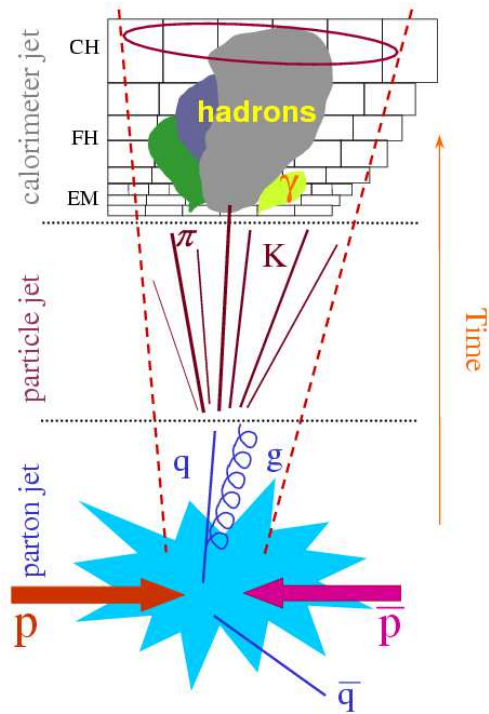}
  \end{center}
  \caption{\it
    Jets clustering can be defined at the parton, particle, and detector levels.
    \label{fig:levels} }
\end{figure}

Figure~\ref{fig:hard_scat} should be thought of as occurring within the
radius of the proton around the colliding partons.  In fact, the
picture becomes more complicated when the property of QCD color
confinement and detector effects are included.  The colored partons
must hadronize into color neutral hadrons.  All of these particles
originating from the different components of the collider event are
indistinguishable in the detector, and it is the job of jet algorithms
to cluster these objects into jets.  Figure~\ref{fig:levels}
illustrates that jets may be clustered at the parton (quarks and
gluons) or particle (hadrons) level when dealing with MC simulation,
or detector (calorimeter towers) levels.  Of course, measurements are
made at the detector level, but it is useful to use the the parton and
particle level jets from MC studies to derive corrections for the
measured quantities.  

Most results discussed in this note will focus on the
properties of the perturbative component of the collision.  However,
studies of the ``underlying event'' ~\cite{UE_runI,UE_runII} from CDF
focus on measuring observables that are sensitive to the
non-perturbative components such as beam remnants and multiple parton
interactions.  These studies provide constraints useful for the
modeling of the non-perturbative regime (where pQCD fails), such as
the ``soft'' interactions generating the underlying event which
accompanies the ``hard'' collision.  

\begin{figure}[htp]
  \begin{center}
      \includegraphics[width=6.5cm]{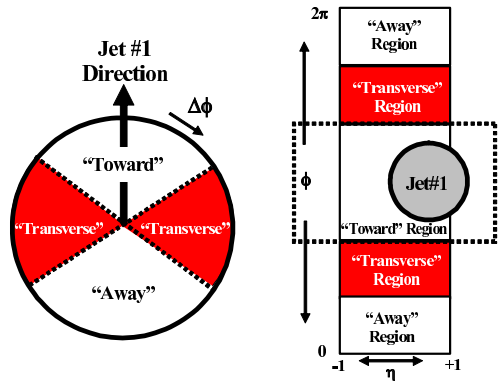}
 \end{center}
  \caption{\it
    Illustration of correlations in azimuthal angle $\phi$ relative to
    the direction of the leading jet in the event.  Observables
    studied in the \TR\ region are sensitive the \UE. 
    \label{fig:UE1} }
\end{figure}

The direction of 
the leading calorimeter jet is used to isolate regions of \etaphi\
space that are sensitive to the underlying event. As 
illustrated in figure~\ref{fig:UE1}, the direction of the leading jet, \Jone, is used to define correlations in the azimuthal 
angle, \delphi.  The angle $\Delta\phi=\phi-\phi_{\rm jet\# 1}$ is the relative azimuthal angle between a 
charged particle (or a calorimeter tower) and the direction of \Jone.
The \TR\ region is perpendicular to the plane of the hard $2$-to-$2$
scattering and is therefore very sensitive to the \UE.  These regions
can be studied for different event topologies such as leading jet
(require one or more jets), back-to-back (requiring two or more jets
with the leading jets back-to-back in $\phi$), and exclusive dijet
(requiring only two jets which are back-to-back in $\phi$).  By
studying different regions and event topologies components of
the hadronic collision can be isolated.

\begin{figure}[htp]
  \begin{center}
    \includegraphics[width=8.5cm]{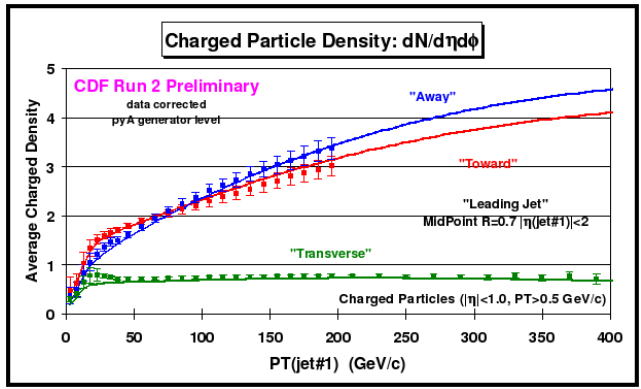}
  \end{center}
  \caption{\it
      The charged particle density per unit $\eta -\phi$ in the
      toward, away, and transverse regions.  The points are the data
      corrected to the particle level and the lines are the \py\
      prediction for each distribution.
    \label{fig:UE2} }
\end{figure}

  CDF has recently updated their UE studies
for leading jet events and other event topologies are under study.  As
an example of the types of observables measured, the charged particle
density per unit $\eta -\phi$ in the toward, away, and transverse
regions is shown in figure~\ref{fig:UE2}.  The goal is to publish more
than one hundred distributions of observables corrected to the particle level.
These results will be useful for tuning and improving theoretical
models of hadronic collisions.  Understanding the underlying event
contribution to jet events is important for many searches at the Large
Hadron Collider (LHC) and measurements of this type will likely be of
the first made at the LHC~\cite{UE_CMS,Skands:2007zz}. 

\section{Jet Cross Section Measurements}

\subsection{\it Inclusive Jet Cross Sections}

The measurement of the differential inclusive jet cross section
at the Tevatron probes the highest momentum transfers in
particle collisions, and thus is potentially sensitive to new physics such as
quark substructure~\cite{NewPhys1}.  The measurement also provides a
fundamental test of predictions of perturbative quantum chromodynamics (pQCD)~\cite{pqcd1,pqcd2}.
Comparisons of the measured cross section with pQCD predictions
provide constraints on the parton distribution function (PDF) of the
(anti)proton, in particular at high momentum fraction ($x \gtrsim 0.3$)
where the gluon distribution is poorly constrained~\cite{CTEQ6.1M}.
Further constraints on the gluon distribution at high $x$
will contribute to reduced uncertainties on theoretical
predictions of many interesting physics processes both for experiments at
the Tevatron and for future experiments at the LHC.  Extending the
measurements to higher rapidities significantly increases the
kinematic reach in the $x$-$Q$ space, where $Q$ denotes the momentum
transfer, and helps to place stronger constraints on the gluon PDF. 

\begin{figure}[htp]
  \begin{center}
\includegraphics[width=6.0cm]{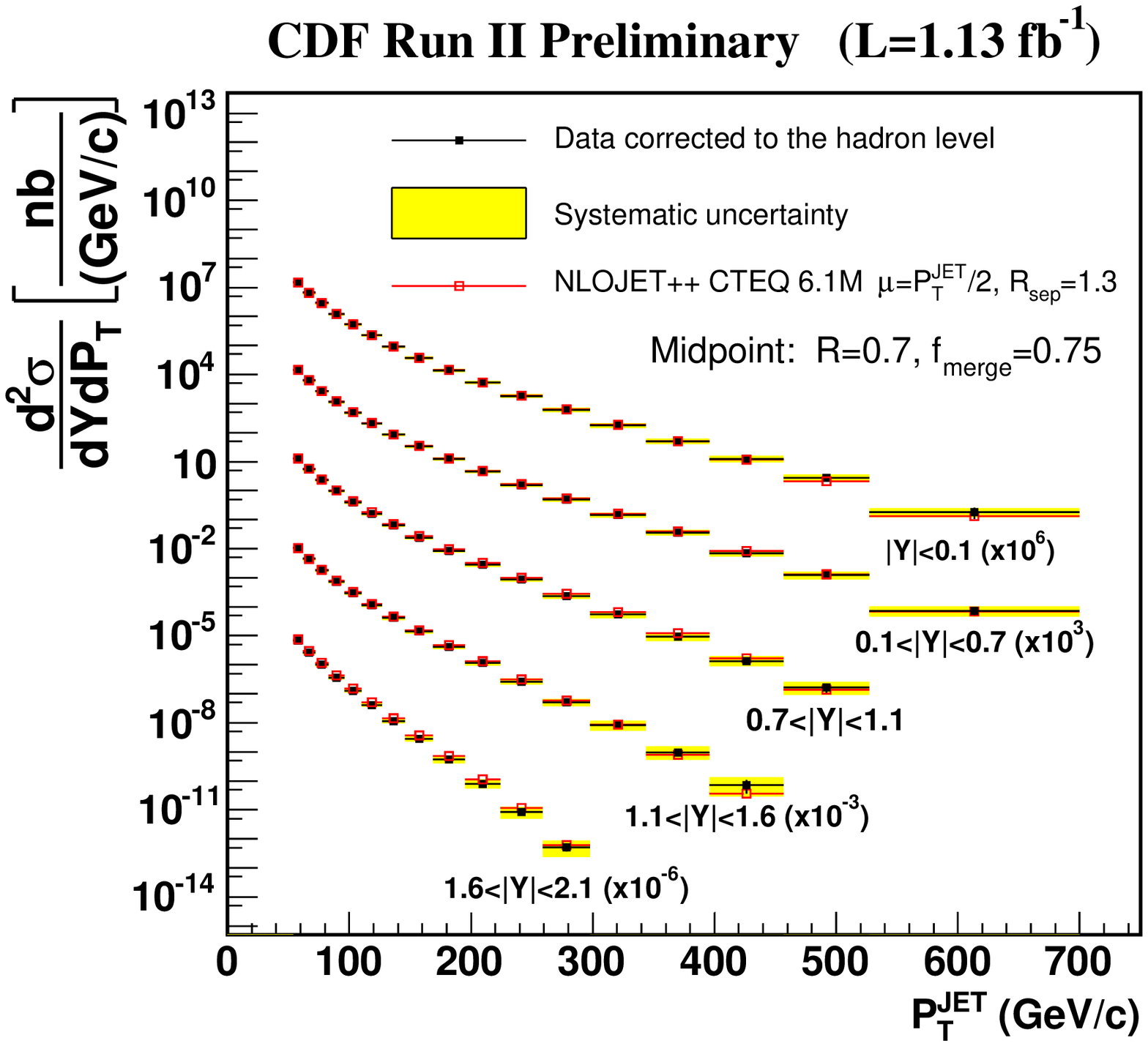}
\includegraphics[width=5.0cm]{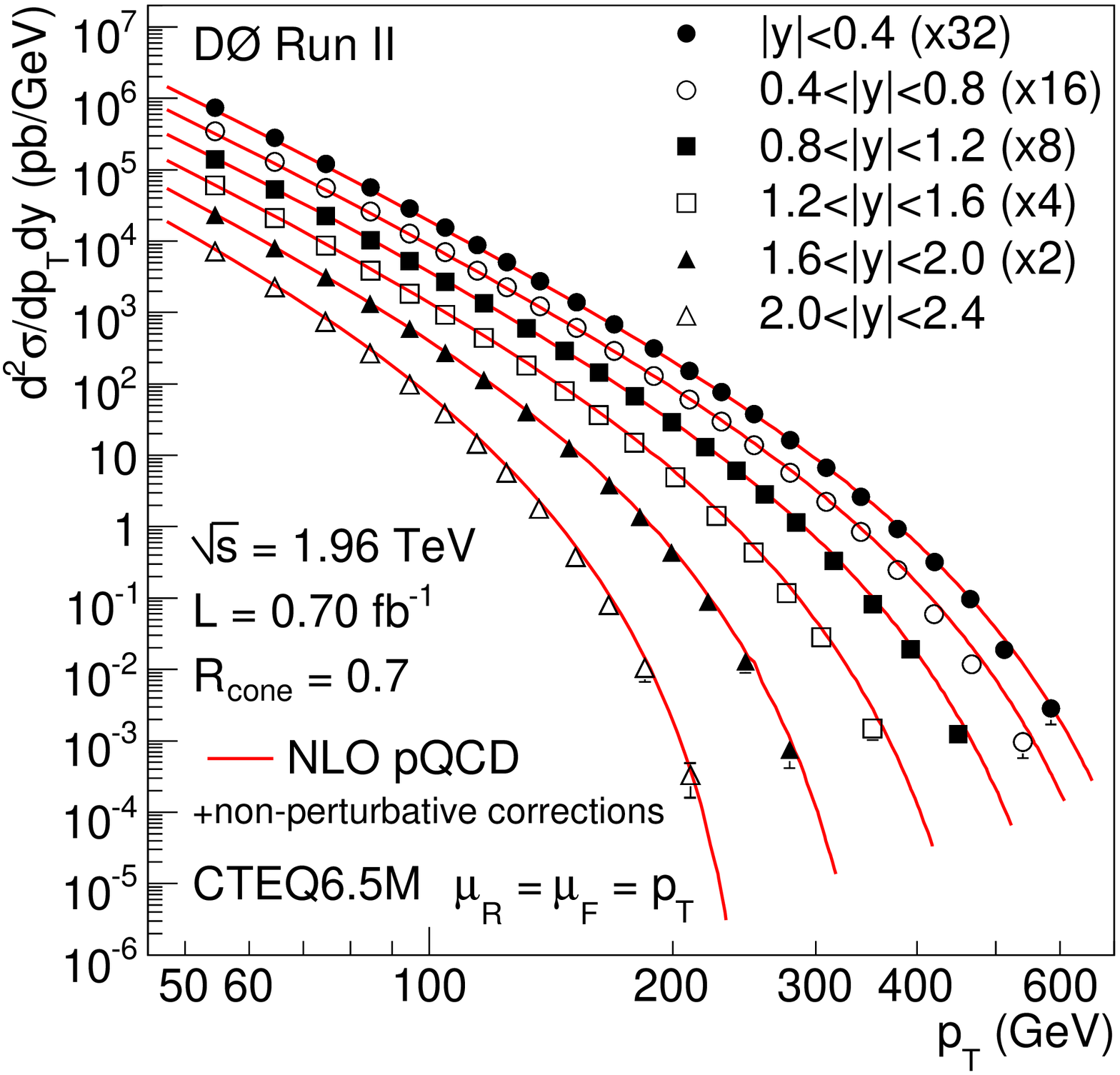}
\end{center}
\caption{\it
    The inclusive jet cross section distributions recently measured by
    CDF (left) and D\O\ (right) using the Midpoint cone-jet clustering
    algorithm.
    \label{fig:IncJetDists} }
\end{figure}

\begin{figure}[htp]
  \begin{center}
\includegraphics[width=9.0cm]{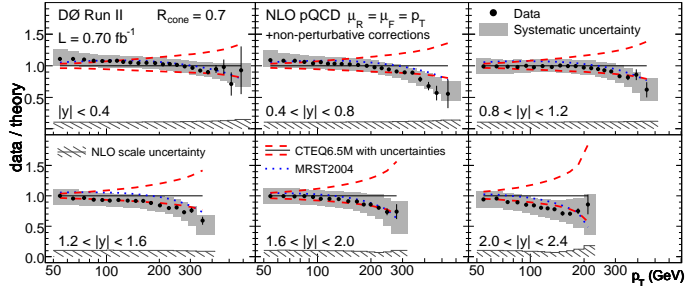}
\end{center}
\caption{\it
    The inclusive jet cross section ratios to the NLO pQCD predictions
    from
    D\O\  using the Midpoint cone-jet clustering
    algorithm.
    \label{fig:IncJetRatiosD0} }
\end{figure}

\begin{figure}[htp]
  \begin{center}
\includegraphics[width=9.0cm]{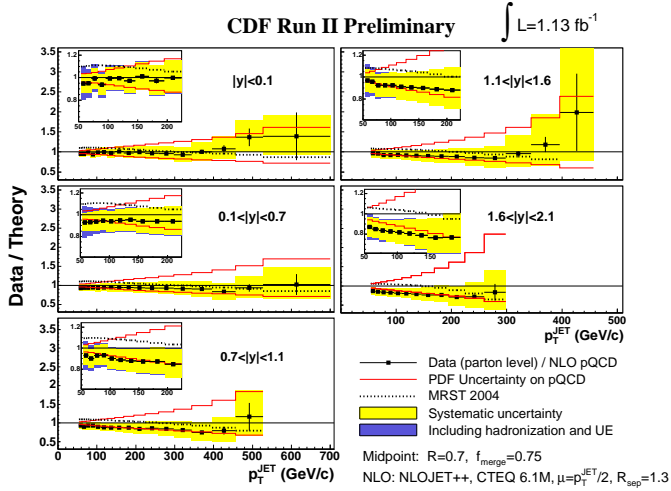}
\end{center}
\caption{\it
    The inclusive jet cross section ratios to the NLO pQCD predictions
    from
    CDF using the Midpoint cone-jet clustering
    algorithm.
    \label{fig:IncJetRatiosCDF} }
\end{figure}

The inclusive jet cross section has been measured in Run
II by CDF ~\cite{Midpoint_RC,KT_PRL} and D\O~\cite{IncJet_D0}.
The most recent measurements using the Midpoint cone jet clustering
algorithm are shown in figure~\ref{fig:IncJetDists}.  The CDF result
(left) compares with NLO predictions using CTEQ6.1M PDFs and
$1.1~\mbox{fb}^{-1}$ of luminosity and breaks the measurement into
five rapidity regions with $|y|<2.1$, while the D\O\ result (right) compares
with CTEQ6.5M~\cite{Nadolsky:2008zw}
using $0.7~\mbox{fb}^{-1}$ of luminosity and splits the rapidity into six
regions with $|y|<2.4$.  The comparison with NLO pQCD is shown by taking
the ratio (DATA/THEORY) in figures~\ref{fig:IncJetRatiosD0}
and~\ref{fig:IncJetRatiosCDF}.  Both measurements observe reasonable
agreement with the NLO predictions and see similar trends in the data
at high rapidities.  In addition the systematic uncertaintiess are smaller than
the PDF uncertainty on the theory prediction and they should therefore be
useful to constrain the proton PDFs.  D\O\ recently reduced their
absolute jet energy scale uncertainty - which yields the dominant
systematic uncertainty in this measurement - to less than 2~\%, and
this improvement will lead to important constraints on the gluon PDF.
These results are also reasonably consistent with the recently
published CDF measurement~\cite{KT_PRD} using the $k_{T}$ clustering
algorithm~\cite{KT} pointing to the conclusion that the $k_{T}$-type algorithm
can work well in the difficult hadron collider environment.

\subsection{\it Dijet Mass}
\begin{figure}[htp]
  \begin{center}
\includegraphics[width=7.5cm]{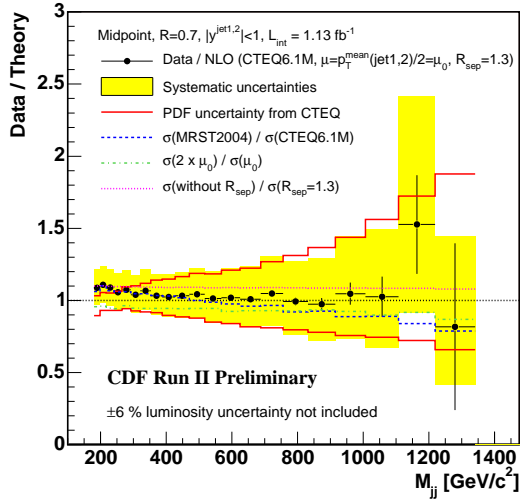}
\end{center}
\caption{\it
    The dijet mass cross section ratio to the NLO pQCD prediction from
    CDF using the Midpoint cone-jet clustering
    algorithm.
    \label{fig:dijet_mass} }
\end{figure}

In addition to being a fundamental test of pQCD
which can be used to constrain PDFs, the dijet mass ($M_{jj}$) cross section
distribution can be used to constrain new physics models which predict
heavy particles decaying to dijets.  A recent measurement from CDF of
the high dijet mass production cross section for $180<M_{jj}<1350$~\gevcc\ uses $1.1~\mbox{fb}^{-1}$ of
luminosity.  As shown in figure~\ref{fig:dijet_mass} nice agreement
with the NLO predictions of NLOJET++~\cite{NLOJET}.

\subsection{\it Exclusive Dijets}
\begin{figure}[htp]
  \begin{center}
    \includegraphics[width=5.5cm]{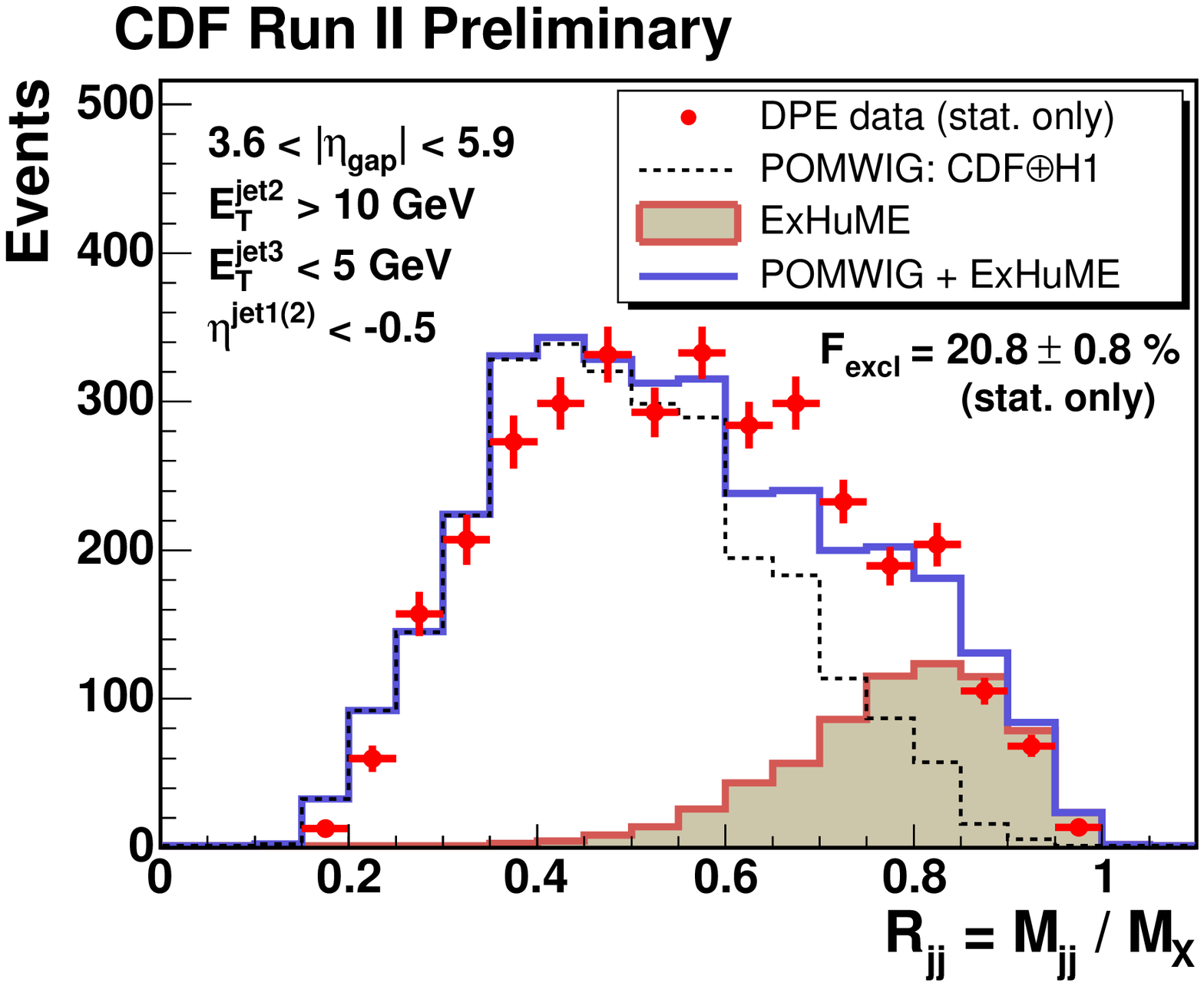}
    \includegraphics[width=5.5cm]{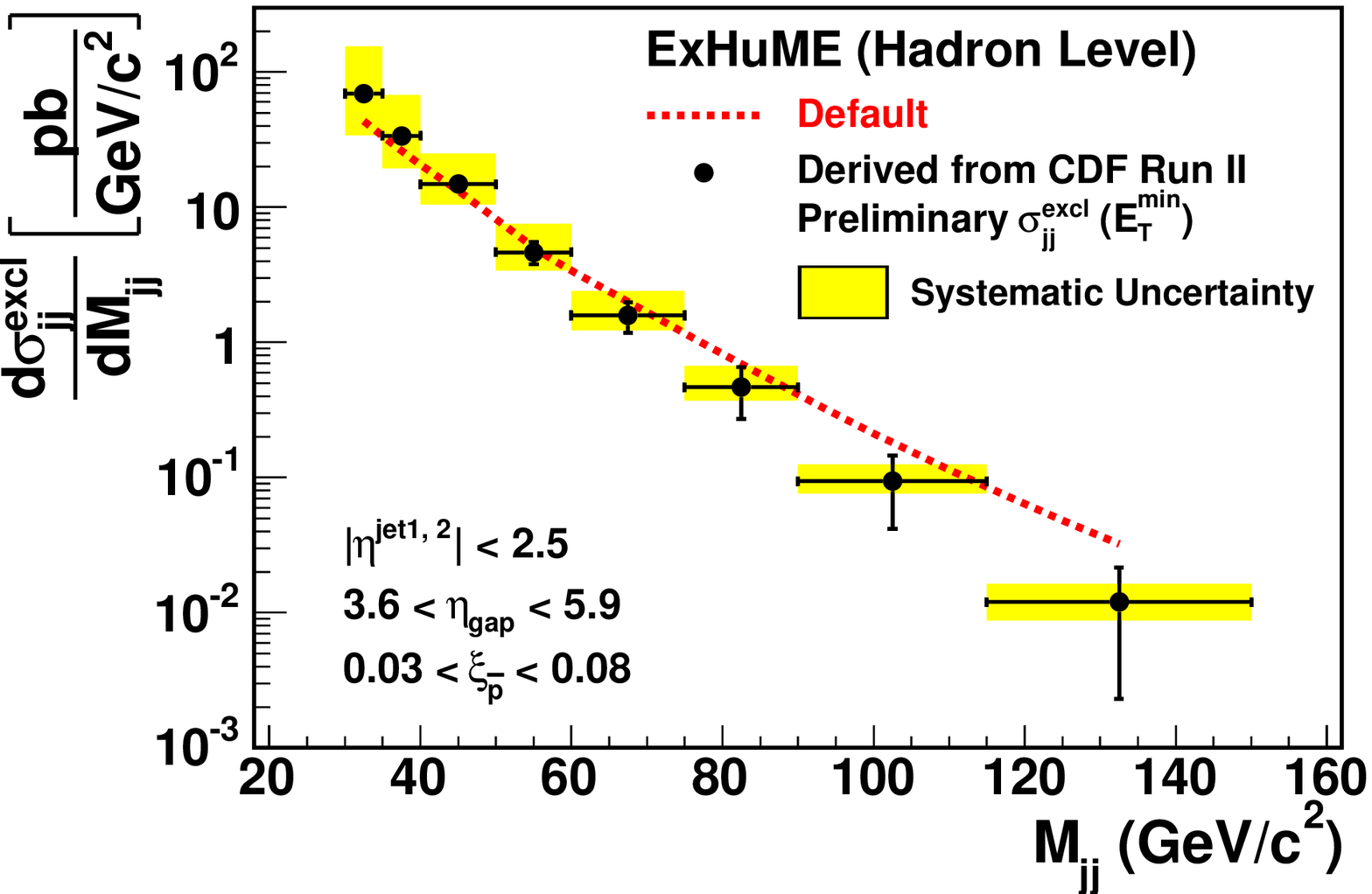}
  \end{center}
\caption{\it
  The Dijet mass fraction (left) and the exclusive dijet mass
  differential cross section distribution (right).
    \label{fig:ExcDijet} }
\end{figure}

In another exciting measurement the first observation of
exclusive dijet production has been reported by CDF~\cite{ExcDijet}.
In this analysis the presence
of exclusively produced dijets ($p+\bar{p} \rightarrow  \bar{p}' +
2jets + p'$) was demonstrated by studying the distributions
of the the dijet mass fraction, defined as the dijet mass divided by
the full system mass.  The dijet mass fraction distributions and the
exclusive dijet mass differential cross section distribution are given
in figure~\ref{fig:ExcDijet}. 

\begin{figure}[htp]
    \begin{center}
      \includegraphics[width=3.0cm]{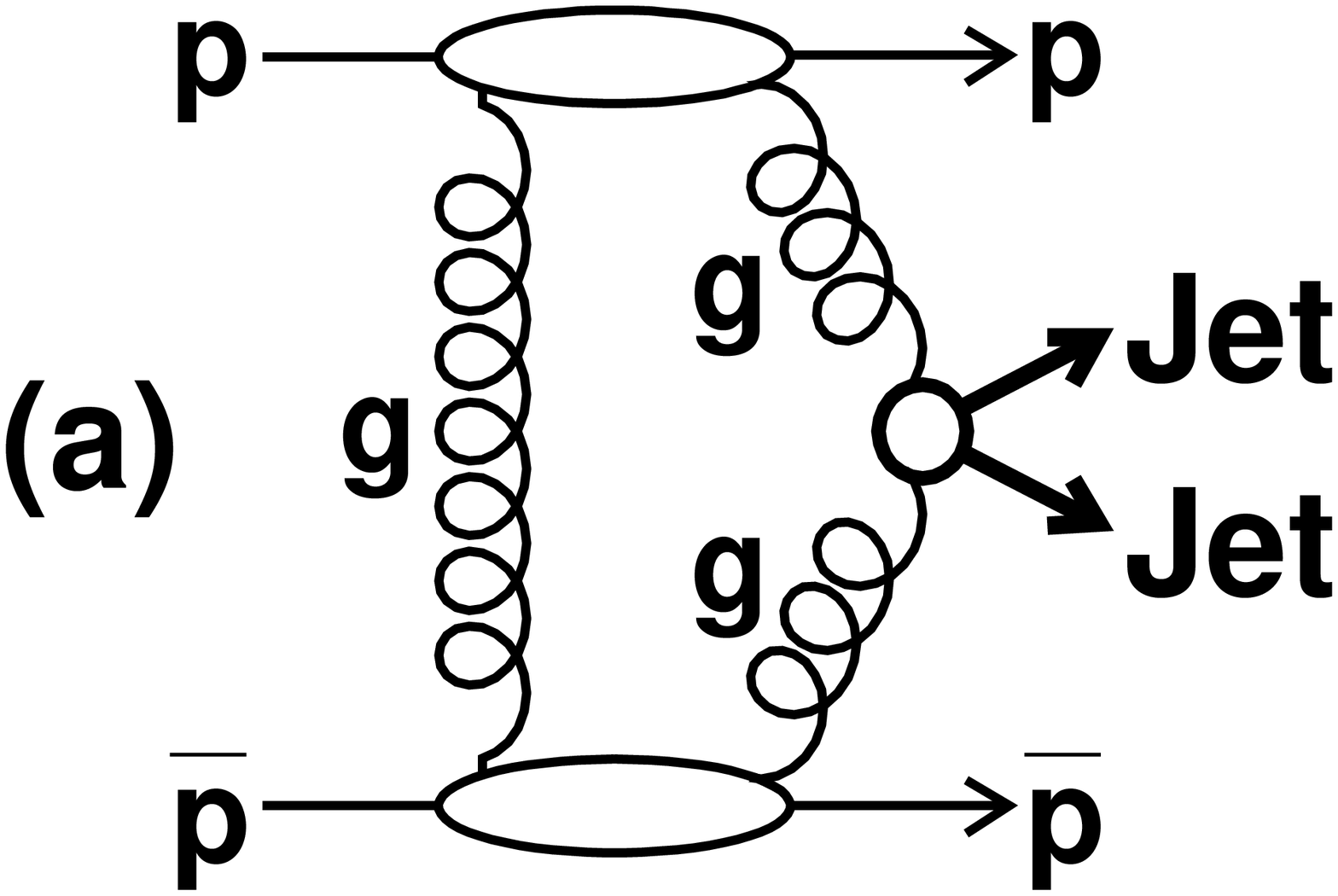}
      \includegraphics[width=3.0cm]{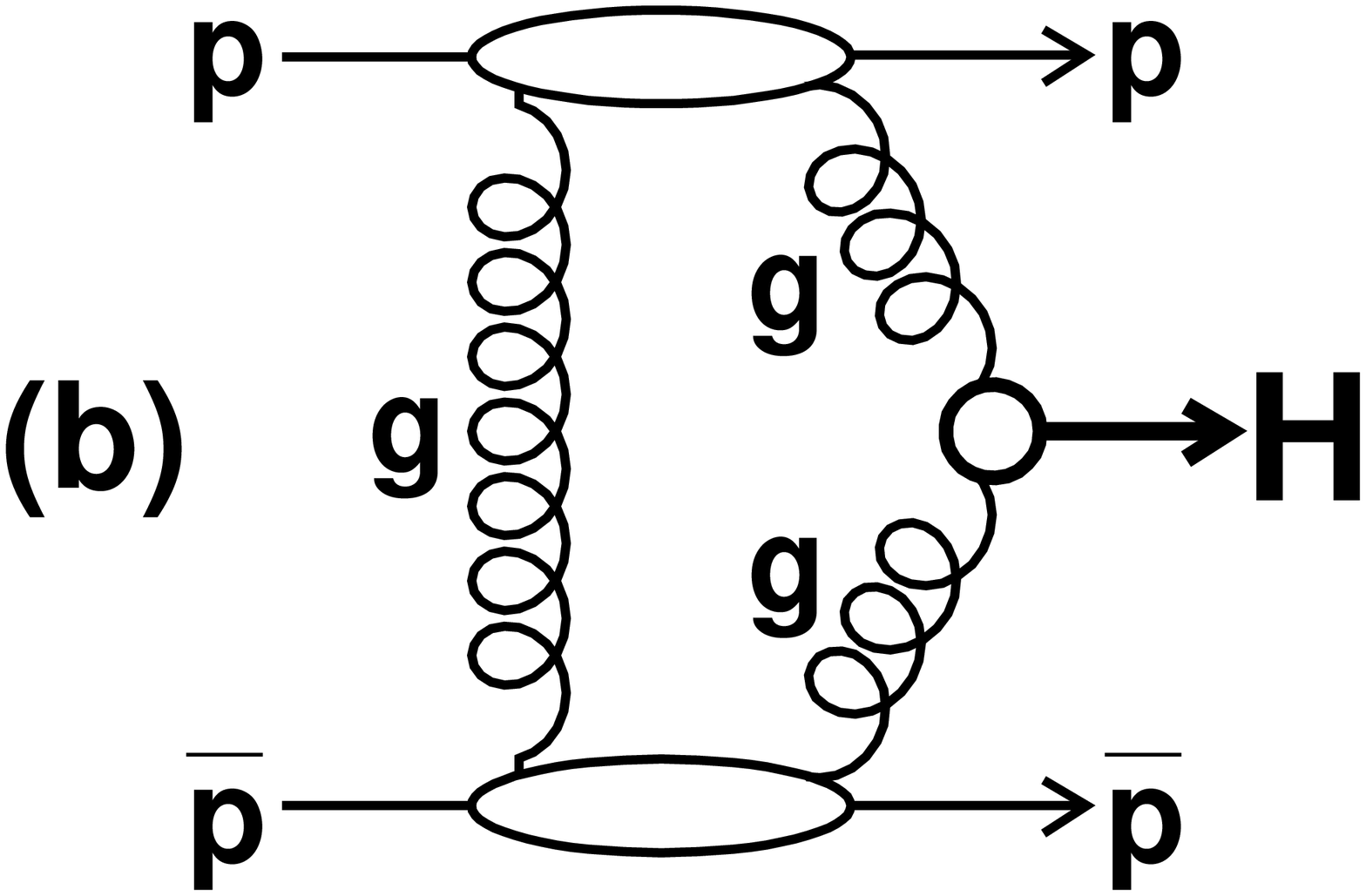}
    \end{center}
    \caption{\it Production diagrams for exclusive dijet (a) and exclusive Higgs
      production (b).
      \label{fig:ExcDiags}}
  \end{figure}

This exclusive dijet result is important because it verifies
that theoretical calculations~\cite{Khoze1,Khoze2} have control over
exclusive production channels like the ones shown in
figure~\ref{fig:ExcDiags}.  The exclusive Higgs boson production
mechanism provides an exciting discovery possibility for the LHC and
this exclusive dijet cross section measurement serves as a useful
calibration channel for this process. 

\section{Boson Plus Jet Measurements}
Boson plus jet production processes measured at the Tevatron
experiments are useful to study pQCD and in addition are some of the
most important backgrounds in new physics searches.  The most recent
results for  $\gamma$ plus jet, $Z$ plus jet, and $W$ plus jet cross
sections are presented next.  

\subsection{\it Triple Differential $\gamma~+ $ Jet Cross Section}
Historically, inclusive direct photon cross section measurements have
observed mediocre agreement with theoretical
predictions~\cite{Alitti:1991yk,Acosta:2002ya,Abazov:2005wc}.
Recently, D\O\ has measured the triple differential $\gamma~+ $ jet
cross section ($\frac{d^{3}\sigma }{dp_{T}^{\gamma}d\eta
  ^{\gamma}d\eta ^{jet}}$) in an effort to understand these
discrepancies ~\cite{Abazov:2008er}.  The analysis requires a photon
in the central region ($|\eta| <1.0$) with $p_{T}>30$~\gevc\ and a jet
in the central ($|\eta| <0.8$) or forward ($1.5<|\eta| <2.5$) region
with $p_{T}>15$~\gevc.  The cross section measurement is then made in
four distinct kinematic regions: 
\begin{itemize}      
\item Region1: Jet and $\gamma$ in the central region and on the same side.
\item Region2: Forward jet and central $\gamma$ in the central region and on the same side.
\item Region3: Jet and $\gamma$ in the central region and on the opposite side.
\item Region4: Forward jet and central $\gamma$ on opposite sides.
\end{itemize}      
\begin{figure}[htp]
    \begin{center}
      \includegraphics[width=8.0cm]{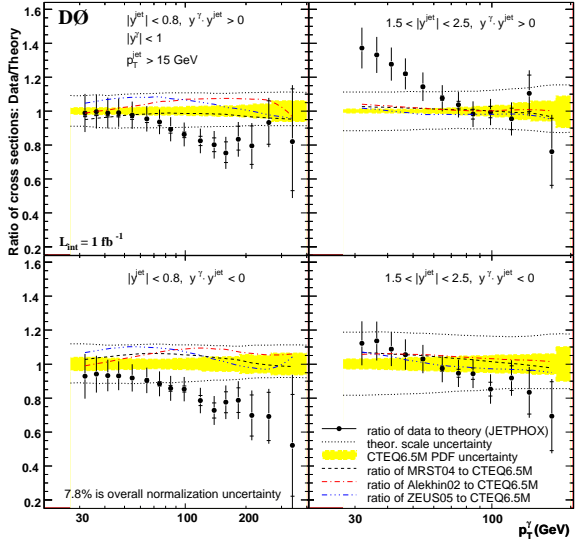}
    \end{center}
    \caption{\it
      Measured cross section to the NLO theory prediction is shown for each kinematic region.
      \label{fig:GammaJet}}
\end{figure}

The ratio of measured cross section to the NLO theory prediction is
shown in figure~\ref{fig:GammaJet}.  This measurement extends the $x$
and $Q$ range significantly over previous measurements.  In many
regions the measured values are outside of the PDF uncertainty bands (CTEQ6.1). In addition, it is clear from the figure that a simple theoretical scale variation cannot bring data and theory into agreement in all four regions.  It should also be noted that the central region results are consistent with previous measurements from UA2, CDF, and D\O.

In addition to the ratios to theoretical predictions, ratios were taken between the different regions.  In these ratios systematic uncertainties on the ratio largely cancel out and total experimental uncertainty is less than 9~\%.  The results of these studies are that shapes are reproduced by theory reasonably well, but there is a quantitative disagreement.

\subsection{\it $Z$ plus Jet Cross Sections}

\begin{figure}[htp]
    \begin{center}
      \includegraphics[width=5.0cm]{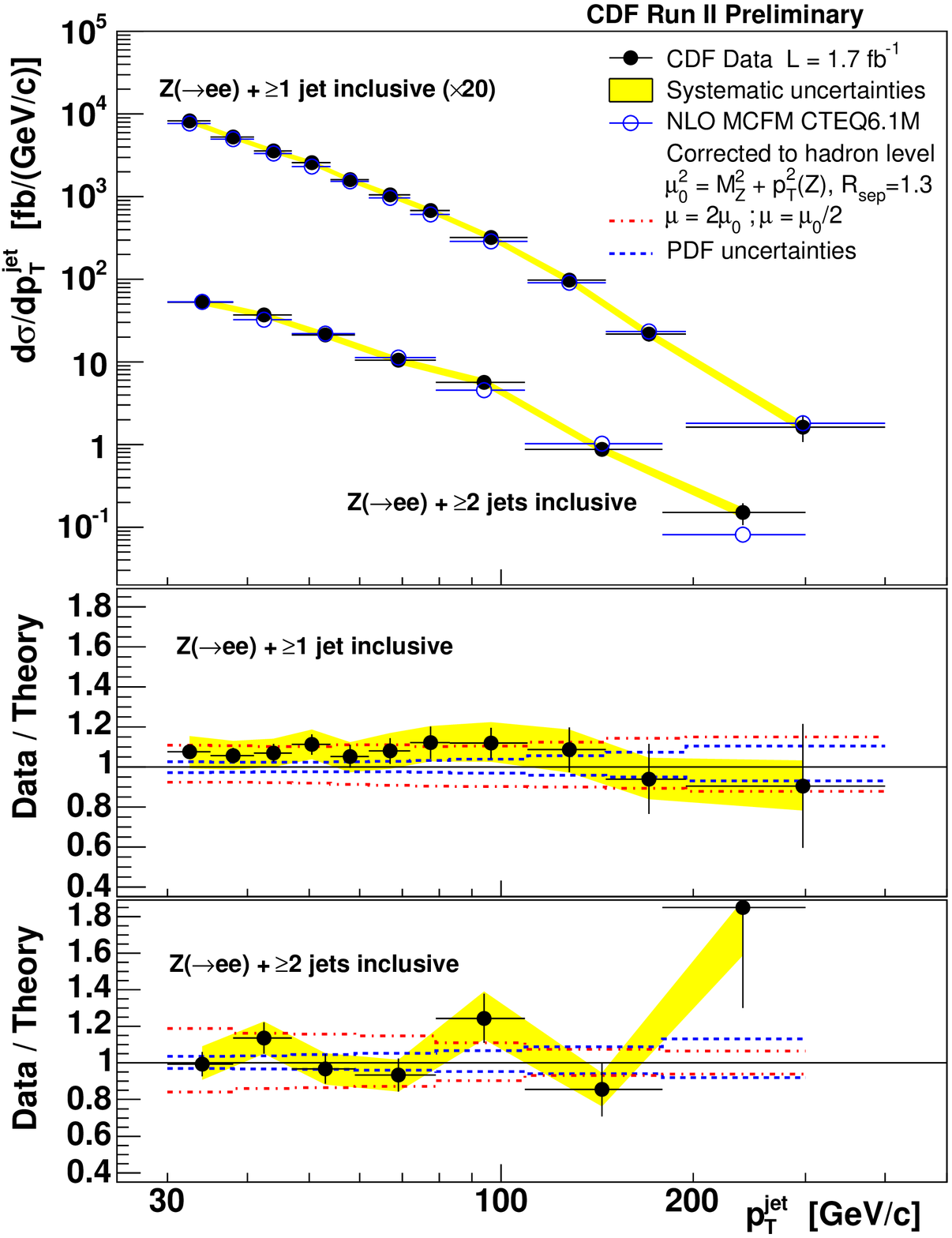} 
      \includegraphics[width=5.0cm]{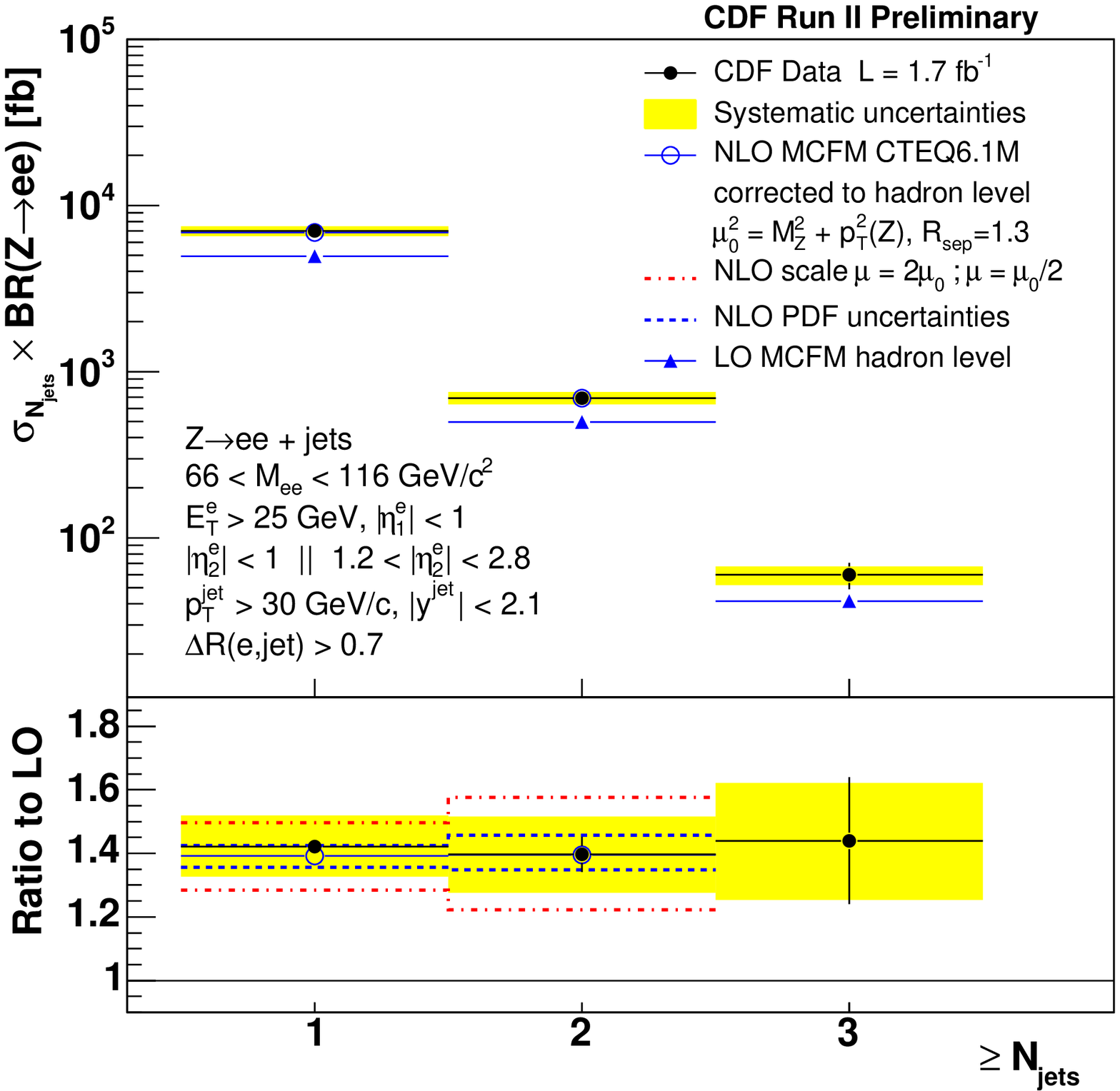}\\
    \end{center}
    \caption{\it
Results of the measurement of the $Z$ plus jet cross section as a function of jet \pt\ (left) and number of jets (right).
      \label{fig:Zjets}}
\end{figure}

$Z$ plus jet production provides a test of the properties of pQCD and this process is the dominant background for many supersymettric searches.  CDF has recently used di-electron final states to measure the inclusive jet cross sections in events with a $Z/\gamma *$~\cite{Zjet:2007cp}.  Figure~\ref{fig:Zjets} shows the jet \pt\ distributions for $\ge 1$ and $\ge 2$ jets (left) and N-jet distributions (right).  Good agreement is observed with the NLO predictions.  The ratio to leading order shown in the N-jet study reveals that the LO-NLO ``k-factor'' does not exhibit strong dependence on the number of jets.

Using di-lepton ($e$ or $\mu$) final states CDF has also recently measured the $Z$ plus $b$-jet cross section.  This measurement probes the heavy flavor content of the proton and is an important background for singly produced top quark, $ZH$, and supersymmetric Higgs searches.  For this analysis the invariant mass distribution for tracks pointing to a displaced vertex is used to separate b, c, and light quark contributions to the $Z$ plus jet events.  In table~\ref{tab:Zbjet} the measured cross section and cross section ratios of $Z$ plus $b$-jet to inclusive $Z$ and $Z$ plus jet cross sections are shown.

\begin{table*}
\tiny
\begin{center}
  \begin{tabular}{|l|c|c|c|c|c|c| } \hline
 & CDF Data & PYTHIA & ALPGEN & NLO \\
 &          &        &        & +U.E+hadr. \\\hline
 $\sigma(Z+ b{\,\rm jet})$ & 0.86 $ \pm 0.14\pm 0.12$   pb   & --   & -- & $
0.53$~pb\\

 $\sigma(Z+ b{\,\rm jet})/\sigma(Z)$  &  0.336 $\pm 0.053 \pm0.041$\%  & 0.35\% 
& 0.21\% & 0.23 \%\\
 $\sigma(Z+ b{\,\rm jet})/\sigma({Z+ \rm jet})$ & 2.11 $\pm 0.33 \pm0.34$\%   & 
2.18\% & 1.45\% & 1.71\% \\ \hline        
\end{tabular}
\end{center}\caption{
The measured cross section and cross section ratios of $Z$ plus $b$-jet to inclusive $Z$ and $Z$ plus jet cross sections as well as theoretical predictions for these quantities from \py, {\sc alpgen}, and {\sc mcfm}~\cite{Campbell:2002tg} with corrections for UE and hadronization affects.
\label{tab:Zbjet}}
\end{table*}

\subsection{\it $W$ plus Jet Cross Sections}

W plus c-jet production is an important background for supersymmetric
top quark and
Higgs production.  In addition the measurement of its cross section
tests the s-quark content of the proton.  Recently D\O\ measured this
cross section and found reasonable agreement with the {\sc alpgen}~\cite{Mangano:2002ea}
prediction~\cite{Abazov:2008qz,Aaltonen:2007dm}.  $W$ plus $b$-jet
production is the dominant background for single top quark and $WH$ searches.  Using a
displaced vertex mass fit, CDF measured the cross section for $W$ plus
$b$-jet events with an electron or muon of $p_{T}>20$~\gevc\ and
$|\eta|<1.1$, missing transverse energy greater than $25$~GeV, and one
or more $b$-tagged jets with $p_{T}>20$~\gevc\ and $|\eta|<2.0$.  The
result is   $\sigma _{W+b-jets}\times Br(W \rightarrow l \nu) = 2.74
\pm 0.27 (stat) \pm 0.42(syst)~pb$.  This measurement should provide
useful constraints to the $W$ boson plus b-jet backgrounds for many future
searches.

\section{QCD Conclusions}

Measurements from the Tevatron Run II are defining a new level of QCD
precision measurements in hadron-hadron collisions.  In this note
many results from the Tevatron's rich program in QCD studies have been
reviewed including: jet cross sections, W+jets, Z+jets, $\gamma$+jets
and more.  The recent inclusive jet cross section measurements from
CDF and D\O\ report nice agreement with NLO predictions and observe
similar trends in the data-theory comparison.  Boson plus jet and
boson plus heavy flavor cross sections are being measured.  These
measurements are important for tests of pQCD and they also provide
important constraints they provide on important backgrounds for new
physics searches for supersymmetry and the Higgs boson.  To summarize, the QCD
programs of the CDF and D\O\ experiments are dedicated to testing and
constraining pQCD and also measuring cross sections of important
background processes.  This important effort will continue to produce
improved results as the Tevatron data sample continues to grow, so
stay tuned.       

  We thank the Fermilab staff and the technical staffs of the
  participating institutions for their vital contributions. This work
  was supported by the U.S. Department of Energy and National Science
  Foundation; the Italian Istituto Nazionale di Fisica Nucleare; the
  Ministry of Education, Culture, Sports, Science and Technology of
  Japan; the Natural Sciences and Engineering Research Council of
  Canada; the National Science Council of the Republic of China; the
  Swiss National Science Foundation; the A.P. Sloan Foundation; the
  Bundesministerium f\"ur Bildung und Forschung, Germany; the Korean
  Science and Engineering Foundation and the Korean Research
  Foundation; the Science and Technology Facilities Council and the
  Royal Society, UK; the Institut National de Physique Nucleaire et
  Physique des Particules/CNRS; the Russian Foundation for Basic
  Research; the Ministerio de Educaci\'{o}n y Ciencia and Programa
  Consolider-Ingenio 2010, Spain; the Slovak R\&D Agency; and the
  Academy of Finland.


\begin{thebibliography}{99}

\bibitem{CDF_detector}
  D.~E.~Acosta {\it et al.} [CDF Collaboration],
  Phys.\ Rev.\  D {\bf 71}, 032001 (2005)
  [arXiv:hep-ex/0412071].

\bibitem{D0_detector}
  T.~LeCompte and H.~T.~Diehl,
  Ann.\ Rev.\ Nucl.\ Part.\ Sci.\  {\bf 50}, 71 (2000).

\bibitem{Sjostrand:2000wi}
  T.~Sjostrand, P.~Eden, C.~Friberg, L.~Lonnblad, G.~Miu, S.~Mrenna and E.~Norrbin,
  Comput.\ Phys.\ Commun.\  {\bf 135}, 238 (2001)
  [arXiv:hep-ph/0010017].

\bibitem{UE_runI}
  T. Affolder, {\it et al} [CDF Collaboration], 
  Phys.\ Rev.\  D {\bf 65}, 092002 (2002).

\bibitem{UE_runII}
  R.~Field and R.~C.~Group  [For The CDF Collaboration],
  arXiv:hep-ph/0510198,~(2005).

\bibitem{UE_CMS}
  D.~Acosta, {\it et al} [The CMS Collaboration],
  CERN-CMS-NOTE-2006-067,~(2007)

\bibitem{Skands:2007zz}
  P.~Z.~Skands, Contribution to Les Houches Workshop on Physics at TeV
  Colliders, FERMILAB-CONF-07-706-T,~(2007)

\bibitem{NewPhys1}
  E.~Eichten, K.~D.~Lane and M.~E.~Peskin,
  Phys.\ Rev.\ Lett.\  {\bf 50}, 811 (1983).

\bibitem{pqcd1}
  D.~J.~Gross and F.~Wilczek,
  Phys.\ Rev.\  D {\bf 8}, 3633 (1973).

\bibitem{pqcd2}
  H.~Fritzsch, M.~Gell-Mann and H.~Leutwyler,
  Phys.\ Lett.\  B {\bf 47}, 365 (1973).

\bibitem{CTEQ6.1M}
  D.~Stump, J.~Huston, J.~Pumplin, W.~K.~Tung, H.~L.~Lai, S.~Kuhlmann and J.~F.~Owens,
  JHEP {\bf 0310}, 046 (2003)
  [arXiv:hep-ph/0303013].

\bibitem{Midpoint_RC}
  A.~Abulencia {\it et al.}  [CDF Collaboration],
  Phys.\ Rev.\  D {\bf 74}, 071103 (2006)
  [arXiv:hep-ex/0512020].

\bibitem{KT_PRL}
  A.~Abulencia {\it et al.}  [CDF Collaboration],
  Phys.\ Rev.\ Lett.\  {\bf 96}, 122001 (2006)
  [arXiv:hep-ex/0512062].

\bibitem{IncJet_D0}
   Abazov, V.M. {\it et al.}  [D0 Collaboration],
  arXiv:0802.2400 [hep-ex].

\bibitem{Nadolsky:2008zw}
  P.~M.~Nadolsky {\it et al.},
  arXiv:0802.0007 [hep-ph].

\bibitem{KT_PRD}
  A.~Abulencia {\it et al.}  [CDF Collaboration],
  Phys.\ Rev.\  D {\bf 74}, 071103 (2006)
  [arXiv:hep-ex/0512020].

\bibitem{KT}
  S.~D.~Ellis and D.~E.~Soper,
  Phys.\ Rev.\  D {\bf 48}, 3160 (1993)
  [arXiv:hep-ph/9305266].

\bibitem{NLOJET}
  Z.~Nagy,
  Phys.\ Rev.\ Lett.\  {\bf 88}, 122003 (2002)
  [arXiv:hep-ph/0110315].

\bibitem{ExcDijet}
  T.~Aaltonen {\it et al.}  [CDF Collaboration],
  arXiv:0712.0604 [hep-ex].

\bibitem{Khoze1}
  V.~A.~Khoze, M.~G.~Ryskin and W.~J.~Stirling,
  arXiv:hep-ph/0607134.

\bibitem{Khoze2}
  V.~A.~Khoze, A.~D.~Martin and M.~G.~Ryskin,
  Eur.\ Phys.\ J.\  C {\bf 14}, 525 (2000)
  [arXiv:hep-ph/0002072].

\bibitem{Alitti:1991yk}
  J.~Alitti {\it et al.}  [UA2 Collaboration],
  Phys.\ Lett.\  B {\bf 263}, 544 (1991).

\bibitem{Acosta:2002ya}
  D.~E.~Acosta {\it et al.}  [CDF Collaboration],
  Phys.\ Rev.\  D {\bf 65}, 112003 (2002)
  [arXiv:hep-ex/0201004].

\bibitem{Abazov:2005wc}
  V.~M.~Abazov {\it et al.}  [D0 Collaboration],
  Phys.\ Lett.\  B {\bf 639}, 151 (2006)
  [Erratum-ibid.\  B {\bf 658}, 285 (2008)]
  [arXiv:hep-ex/0511054].

\bibitem{Abazov:2008er}
  V.~M.~Abazov {\it et al.}  [D0 Collaboration],
  arXiv:0804.1107 [hep-ex].

\bibitem{Zjet:2007cp}
  T.~Aaltonen {\it et al.}  [CDF Collaboration],
  Phys.\ Rev.\ Lett.\  {\bf 100}, 102001 (2008)
  [arXiv:0711.3717 [hep-ex]].

\bibitem{Campbell:2002tg}
  J.~Campbell and R.~K.~Ellis,
  Phys.\ Rev.\  D {\bf 65}, 113007 (2002)
  [arXiv:hep-ph/0202176].

\bibitem{Mangano:2002ea}
  M.~L.~Mangano, M.~Moretti, F.~Piccinini, R.~Pittau and A.~D.~Polosa,
  JHEP {\bf 0307}, 001 (2003)
  [arXiv:hep-ph/0206293].

\bibitem{Abazov:2008qz}
  V.~M.~Abazov {\it et al.}  [D0 Collaboration],
  arXiv:0803.2259 [hep-ex].

\bibitem{Aaltonen:2007dm}
  T.~Aaltonen {\it et al.}  [CDF Collaboration],
  Phys.\ Rev.\ Lett.\  {\bf 100}, 091803 (2008)
  [arXiv:0711.2901 [hep-ex]].



\end{thebibliography}
\end{document}